\documentclass[conference]{IEEEtran}

\usepackage{cite}
\usepackage{amsmath,amssymb,amsfonts}
\usepackage{algorithmic}
\usepackage{graphicx}
\usepackage{textcomp}
\usepackage[table]{xcolor}
\usepackage{float}
\def\BibTeX{{\rm B\kern-.05em{\sc i\kern-.025em b}\kern-.08em
    T\kern-.1667em\lower.7ex\hbox{E}\kern-.125emX}}

\usepackage[hidelinks]{hyperref}

\begin{document}

\title{L2 Ethernet Switch VLSI Implementation}

\author{
    \IEEEauthorblockN{Aniruddh~Mishra}
    \IEEEauthorblockA{\textit{Chandra Department of ECE} \\
    \textit{University of Texas at Austin}\\
    Austin, TX \\
    aniruddh@utexas.edu}
    \and
    \IEEEauthorblockN{Benjamin Oommen}
    \IEEEauthorblockA{\textit{Department of Computer Science} \\
    \textit{University of Texas at Austin}\\
    Austin, TX \\
    baoommen@utexas.edu}
    \and
    \IEEEauthorblockN{Jimmy Liang}
    \IEEEauthorblockA{\textit{Chandra Department of ECE} \\
    \textit{University of Texas at Austin}\\
    Austin, TX \\
    jimmy.liang@utexas.edu}
    
    \thanks{Manuscript received December 11, 2025}
}

\IEEEoverridecommandlockouts

\maketitle

\begin{abstract}
Ethernet switches are foundational to the global internet infrastructure. These devices route packets of data on a local area network between source addresses to destination media access control addresses. On the L2 layer of the Open Systems Interconnections model, Ethernet switches take in digitized data from a Media Independent Interface and send it to the corresponding output port for the destination address. Switches need to handle parallel input and output streams from each port, prioritizing throughput, efficiency, and packet integrity. Due to the confidential nature of the networking device industry, there do not exist many open source implementations of switching fabrics. We propose an open source design for an L2 Ethernet switch along with the power, performance, and area tradeoffs for architecture decisions.
\end{abstract}


\section{Introduction}

\IEEEPARstart{T}{his} 
project aims to implement a Layer 2 Ethernet network switch that complies with the IEEE 802.3 Ethernet standard. The switch will interact with the outside world by receiving and transacting using the PHY layer through a Media Independent Interface (MII). Specifically, this project focuses on the GMII (Gigabit MII) protocol for its simplicity, gigabit speed, and timing margins. The switch should efficiently route incoming packets from their source address to the intended destination ports. A port can have any number of addresses, so the switch must learn and store the addresses, while also evicting when exceeding the maximum number of stored addresses. We chose to have each port duplex as this is standard in modern performance-driven network switches in order to eliminate data collisions and maximize bandwidth. The switch will demonstrate all aspects of core L2 forwarding and routing functionality, including incoming data parsing, MAC address learning, memory access arbitration, buffering/scheduling, and output transmitting. 

This implementation aims to meet all functionality and signal integrity standards for Ethernet as well as maximizing throughput with an awareness of physical design tradeoffs. To maximize correctness, the switch implements a functional scheduling algorithm and interfaces with all ports to get data forwarded to its correct destination. This is done through ensuring no packet corruption, all data sent to their respective correct ports, and proper MAC learning behavior with given source and destination addresses. The switch also handles multiple simultaneous writes and can handle gigabit-speed traffic without packet drops and predictable low latency. This means enabling high performance on multiple ports simultaneously, low latency across the entire switch, and monitoring for buffer sizing/shared memory utilization.

The architecture of this switch is broken down into many different modules. The overall dataflow begins from the PHY's GMII signals, travels in a port into the RX module, accesses memory through memory write side, goes through address learning, leaves using memory read side and the output virtual output queues (VOQs), and finally moves through the TX module back to the PHY's GMII signals. 

\begin{figure} [H]
    \centering
    \includegraphics[width=1\linewidth]{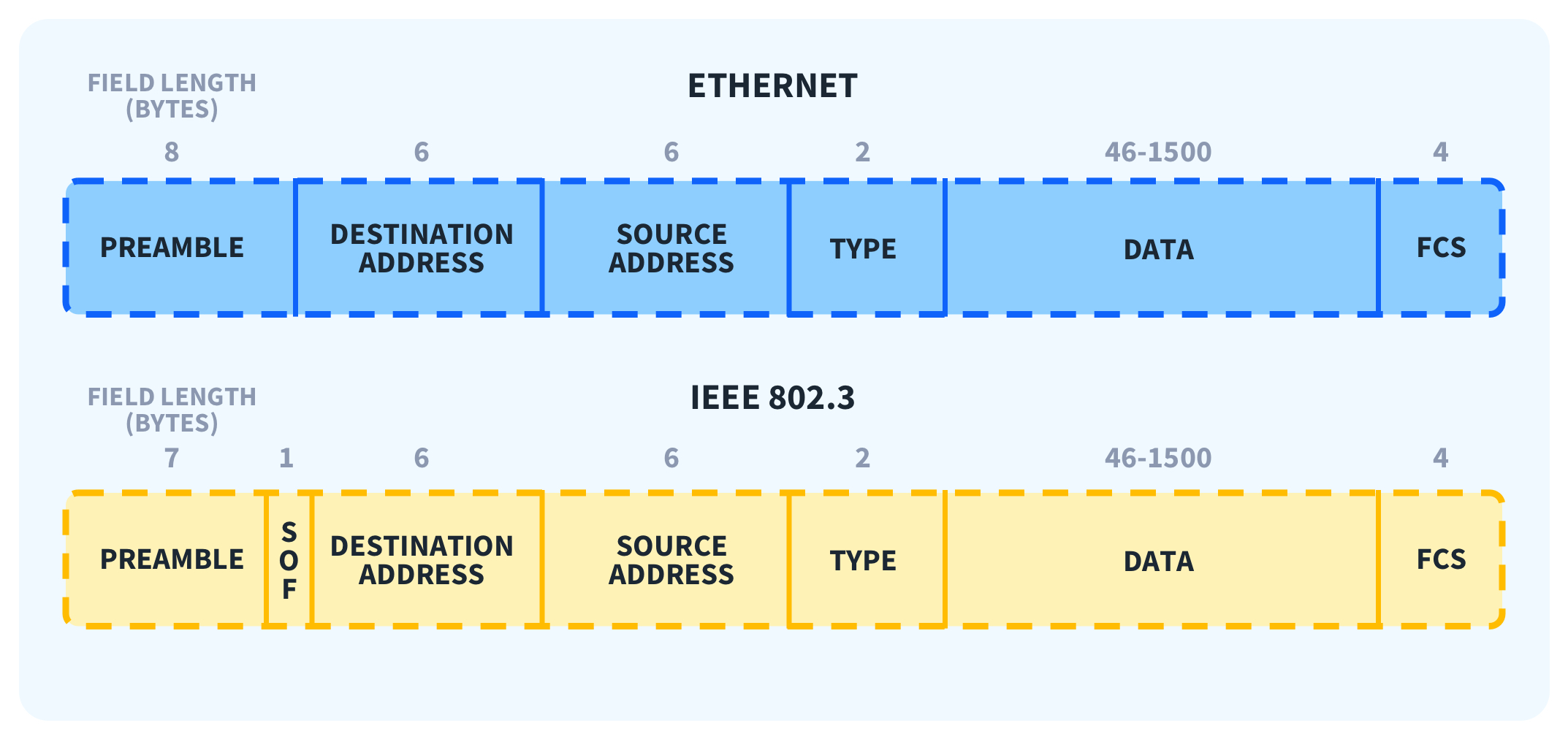}
    \caption{Ethernet Frame Diagram \cite{ethernet_frame}}
    \label{fig:ethernet_frame}
\end{figure}

The Ethernet protocol shown in figure \ref{fig:ethernet_frame} specifies that every frame of data must have the following:
The preamble allows for synchronization of the switch to the PHY layer, represented as a sequence of alternating 1s and 0s. Included in the preamble (or considered separately) is the Start Frame Delimiter (SFD). This is the 8th byte following 7 bytes of the preamble and is just a specific bit sequence (10101011) that shows the beginning of the frame.
The destination and source address is contained as the first part of the actual data in the frame. The source address helps the network switch learn which ports contain which given MAC addresses, while the destination tells the switch where the packet of data should be routed to.
The EtherType of the frame is based on the specific Ethernet data protocol being used. Usually, this part of the frame contains information on the length of the data. Since our network switch is impartial to Ethernet protocol formats, we will treat the type segment as a part of additional data.
The data/payload includes the actual information to deliver to the destination and can vary in size.
The Frame Check Sequence (FCS) is 4 bytes and is calculated using a CRC32 (Cyclic Redundancy Check) 32-bit checksum algorithm. This calculation is done from the input device and the receiving switch in order to detect errors in the data transmitted.

There are other successful techniques to implement an Ethernet network switch. It is possible to build a high-throughput switch that uses per port VOQs and a dynamic scheduling algorithm. Tiny Tera's switch core is a stack of 1-bit crossbar slices connected with high speed serial links. The descriptor based queuing strategy also helps for reliable, fast interactions with memory, which will be drawn upon in our memory arbitration strategy \cite{tiny_tera}.

Many techniques to create fast shared memory switches also exist, where the goal is to construst a high speed network switch out of memroy blocks that individually operate slower than the line rate, or maximum raw speed excluding overheads. The paper presents a fast centralized memory, showing ways to surpass the throughput of a single memory block for a faster network switch overall \cite{fast_shared_memory}.

For implementation details, the bulk of the work on this project focuses on the RTL level HDL code, written in SystemVerilog. All SystemVerilog modules were written, debugged, and verified by us, and the details of the implementation can be found at our GitHub page\footnote{\url{https://github.com/aniruddh-mishra/Ethernet_Network_Switch}}. 

\section{Overall System Architecture} \label{overall_arch}


The goal of the overall architecture is to connect input port data to output signals. Since the Ethernet switch allows for parallel transmission, the system architecture needs to allow connections between each ingress module with each egress module. Additionally, a notable feature of the Ethernet switch is to route the data to the correct destination port based on the specified MAC address. However, when the port first turns on, it will not recognize any destination addresses. To handle this, the switch will flood inputs into all outputs when the routing is unknown. This means that the overall architecture needs to handle a overall flood case, in which all inputs flood to all outputs. Since the Ethernet switch uses 4 ports, this means that 4 packets could simultaneously be sent to all 4 outputs. 

With the former requirements noted, it is clear that the data needs to be queued into output ports because input data can come in faster during flood cases. Many modern implementations, including ours, have a centralized memory with decentralized output queues. Rather than copying the frames into the output queue of each egress module, the virtual output queue implementation means that the frame is only written into one shared memory unit. Each virtual output queue instead stores pointers to the frame data in memory.

\begin{figure} [H]
    \centering
    \includegraphics[width=1\linewidth]{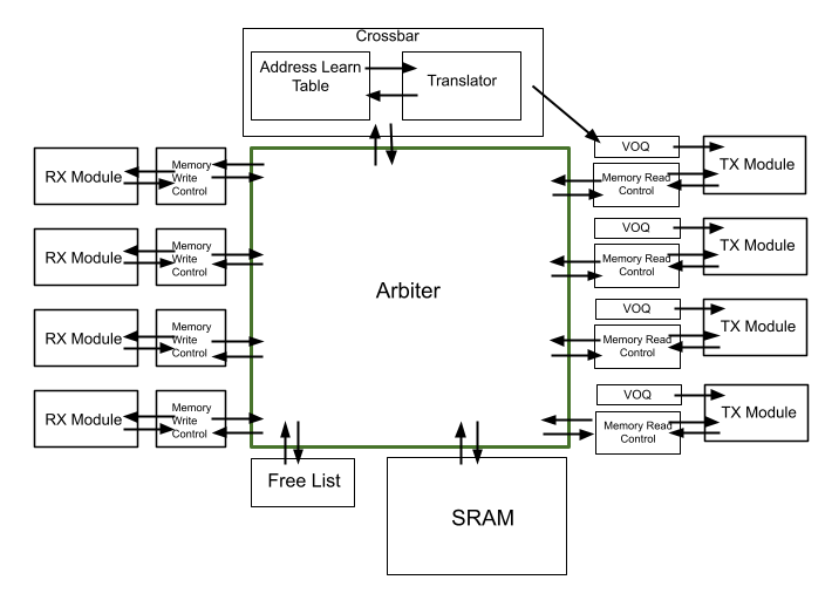}
    \caption{Overall Top Architecture Diagram}
    \label{fig:architecture_diagram}
\end{figure}

\begin{figure} [H]
    \centering
    \includegraphics[width=0.9\linewidth]{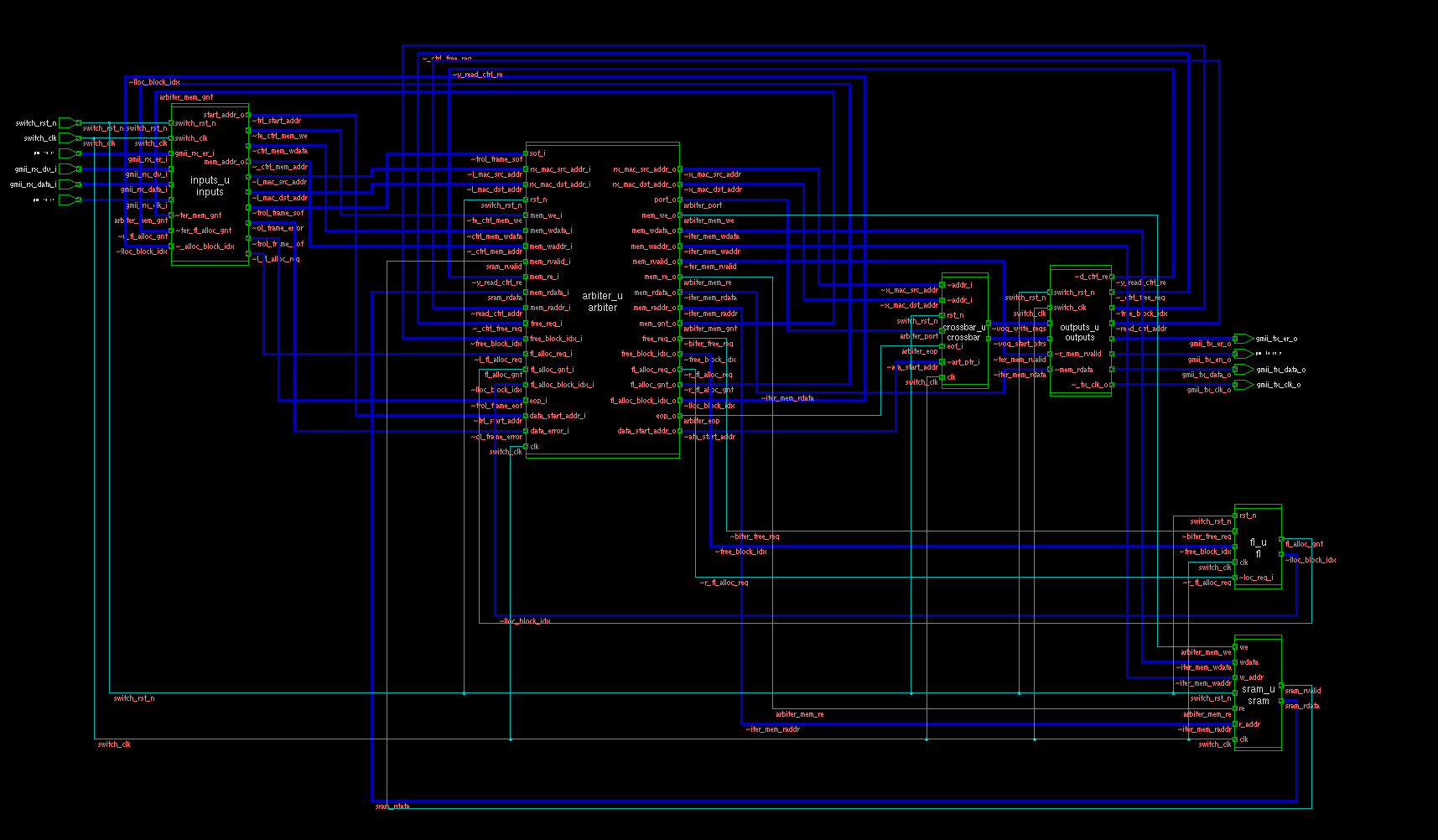}
    \caption{Top level schematic}
    \label{fig:top}
\end{figure}

Another top level architecture decision was arbitrating the memory write ports. The two solutions for this were to make multiple write ports for the SRAM or to use a round robin arbitration scheme to allow a single ingress port access to memory each clock cycle. If we chose to do parallel writes, the memory could either be divided into blocks per ingress port or utilize a complex allocation scheme, which provided a write index to each ingress port. In the case of multiple memory modules, we would be assuming that each ingress port would be transmitting an equivalent amount. On the other hand, the complex allocation would require four write ports in the SRAM, which would increase area and power utilization of the memory. Since the memory is the largest block in the Ethernet switch, we decided to move on with a round robin arbitration scheme. These choices are illustrated in figure \ref{fig:architecture_diagram} and figure \ref{fig:top}.

The overall IO architecture is simply the input and output GMII interfaces. These are all described in table \ref{tab:io_description}. All interface ports except for the switch clock and reset signal are arrays. This is because each port gets an individual GMII connection to the PHY layer.

\begin{table}[H] \scriptsize
    \caption{IO Description of Overall Architecture}
    \centering
    \begin{tabular}{|c|c|>{\raggedright\arraybackslash}p{3cm}|}
        \hline
         \rowcolor{lightgray} \textbf{Pin Name} & \textbf{IO Mode} & \centering\arraybackslash\textbf{Description} \\
         \hline
         Switch\_Clk & Input & The hardware of the internal switch modules between the rx and tx blocks run on the frequency specified by switch\_clk (~500 MHz). \\
         \hline
         Switch\_Rst\_N & Input & The reset signal for the switch is asynchronous negative edge assert and synchronous deassert. \\
         \hline
         GMII\_RX\_CLK & Input & GMII protocol signal that specifies data clock for each ingress port (~125 MHz). \\
         \hline
         GMII\_RX\_DATA & Input & GMII protocol signal that specifies parallel byte data synched with GMII\_RX\_CLK for that specific module \\
         \hline
         GMII\_RX\_DV & Input & GMII protocol signal that specifies data valid for the GMII\_RX\_DATA signal. \\
         \hline
         GMII\_RX\_ER & Input & GMII protocol signal that indicates error with the incoming data frame. \\
         \hline
         GMII\_TX\_CLK & Output & GMII protocol signal that sets the output data clock (~125 MHz). \\
         \hline
         GMII\_TX\_EN & Output & GMII protocol signal from the egress port that enables the valid for the next receiver. \\
         \hline
         GMII\_TX\_ER & Output & GMII protocol signal that indicates error with the output data. \\
         \hline
         GMII\_TX\_DATA & Output & GMII protocol signal that specifies the output data from egress port. \\
         \hline
    \end{tabular}
    \label{tab:io_description}
\end{table}

\section{Ingress Port}

At each port connected to the switch, there is an ingress (input/receiver side) and an egress (output/transceiver side). We chose to use the Gigabit Media Independent Interface (GMII) protocol, which is a standard parallel interface in Ethernet that connects the switch to the PHY for gigabit speeds (1 Gbps). It provides a simple, media-independent way for the network switch to send and receive data to/from the PHY, using an 8-bit parallel bus at 125 MHz. In general, there are multiple different Media Independent Interface (MII) options that are IEEE compliant and are used to interface between the MAC network switch and the PHY. GMII is used over other options when simplicity, gigabit speed, and timing margin are important. GMII has higher throughput than classic MII, has more flexible timing than RMII/RGMII (reduced/reduced gigabit), and is simpler to validate compared to SGMII (serial) which is more suited for signal integrity on longer traces.

\begin{figure}[H]
    \centering
    \includegraphics[width=1\linewidth]{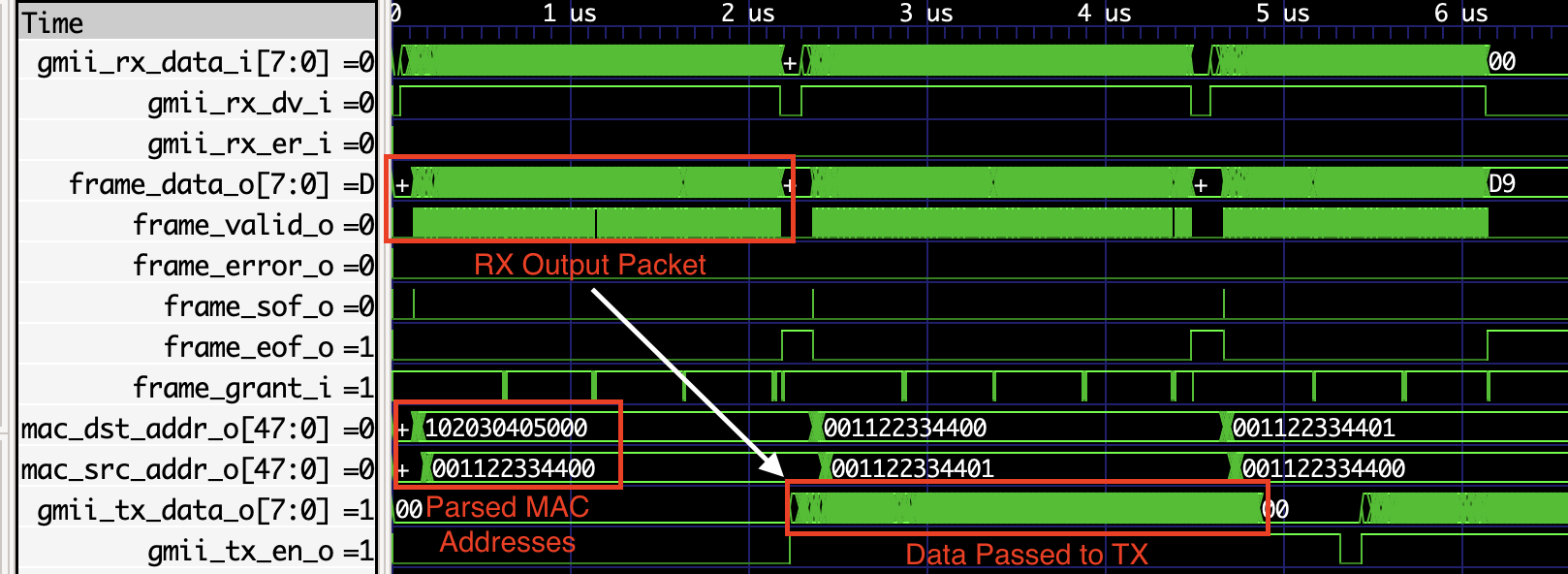}
    \caption{RX Parsing and Sending Packet to Memory}
    \label{fig:rx_mac_control}
\end{figure}

The RX module serves two main functions. It must parse the received Ethernet frame for the MAC destination and source addresses to the crossbar/address table, and it must prepare the Ethernet frame's data for the per-port memory write controllers to access the switch's memory. In figure \ref{fig:rx_mac_control}, data first enters on the GMII RX data input. The RX module forwards this data to memory per-byte on the frame data output. Simultaneously, the destination and source addresses are saved and outputted for address learning. Eventually, this data is confirmed to be transmitted on the TX egress port closely after the frame ends transmission in RX. This shows minimal delay and buffering between first knowing the frame is valid (which can only occur on the last CRC check) and the data beginning to transmit.

For inputs, the RX module feeds the GMII one byte-wide data input into an instantiation of the asynchronous FIFO module in order to cross though the Clock Domain Crossing (CRC) boundary. The asynchronous FIFO handles any metastability that could occur due to a difference in the GMII clock vs switch clock. This metastabilty could occur due to real-world differences in frequency and phase betweeen the clocks. GMII data valid and error signals must similarly cross this CDC boundary, which is handled by using a 2-stage FF synchronizer. Generally, asynchronous FIFOs are necessary for any application that handles multiple CDC bits at a time since metastability issues can cause individual bits to be out of sync with other parallel bits. However, it is acceptable for these values to be one cycle off in comparison to each other or to GMII data since this will be handled by the logic in the 4x faster switch clock domain.

Also, the switch's asynchronous reset signal must be synced to the FIFO's write side domain since the de-assertion of reset being in a different clock domain than the GMII could cause metastability. Thus, a synchronizer module is used which allows for asynchronous assertion but uses a 2 FF synchronizer for synchronous de-assertion.

When the RX module is idling, it checks for the preamble by looking for seven preamble bytes (8'h55) followed by a start frame delimiter (SFD) byte (8'hD5). If the preamble/header counter successfully increments to 8, then this is a valid frame that must begin getting parsed. The destination, source, and type then correspond respectively to subsequent counter values based on their respective lengths (i.e. counter values 8-13, 14-20, and 20-22).

\begin{figure}[H]
    \centering
    \includegraphics[width=1\linewidth]{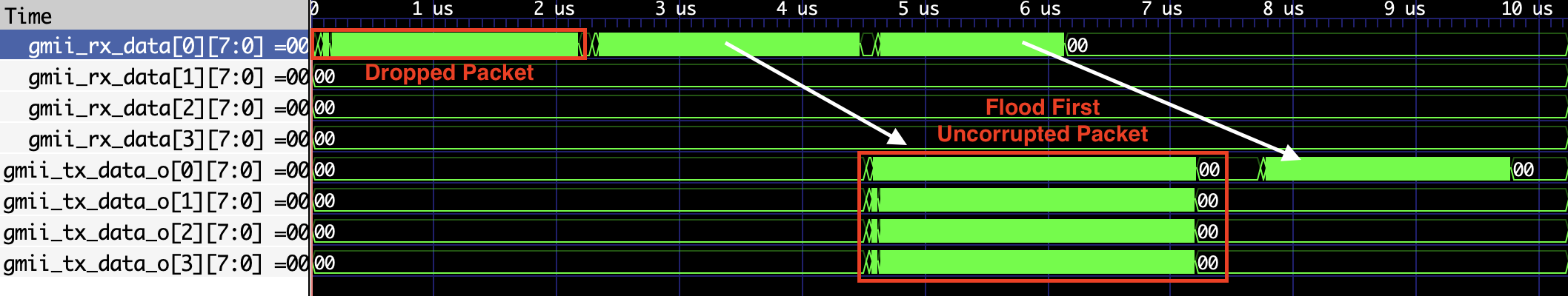}
    \caption{Packet drop functionality with corrupted CRC}
    \label{fig:packet_drop}
\end{figure}

During this time, the RX module reads from the asynchronous FIFO any time the FIFO is not empty. Every byte of data read gets outputted to the memory write controllers along with a high frame valid bit. The first valid data also has a start of frame (SOF) bit set to indicate to memory that a new frame is beginning. Since the switch clock is much faster than the GMII clock, there will always be times the FIFO is empty between valid bytes of output data. This valid ensures that only correct data is processed by the memory write controller. Meanwhile, CRC register continuously updates with every new byte of valid data. The RX module continues sending data and updating its CRC until the switch clock synced GMII data valid goes low, signaling the end of the data frame. Once the last byte of data is sent out, the module needs to handle that the FCS consists of the last four bytes of the frame and should not be included in the CRC calculation. The RX module now uses its CRC and data buffer to check the CRC before the FCS against the FCS. If there is a mismatch, the error is flagged to memory on the same cycle as the end of frame (EOF) bit, as shown in figure \ref{fig:packet_drop}. Notice that the data in the figure does not appear on the TX transmission side, showing that the packet is successfully dropped by the switch due to the store and forward architecture. These values are kept high until the next frame/SOF so the memory read controller can properly handle the end of each frame.

If too much back pressure occurs from memory, the RX module will begin dropping bytes of data, creating an invalid frame. This causes a mismatch in the CRC, thus properly flagging an error with the EOF and resulting in a similar result as figure \ref{fig:packet_drop}. Also, there is no PAUSE handshaking protocol between the sender and the switch, the higher level TCP (Transmission Control Protocol) will handle any resending of frames necessary. 

Initially, this module contained logic in both the GMII and switch clock domains. This meant having two separate state machines and more logic on both the read and write side of the asynchronous FIFO. By moving the logic all solely to the switch clock domain, the write side of the asynchronous FIFO is much simpler, reducing area/power/timing. This reduced the size of the file by 55 percent.

During times when memory is backlogged and memory grant is de-asserted, RX freezes the majority of its sequential logic to minimize unnecessary power usage. The module must continue to check for data valid to go low to keep track of if a frame has ended transmission. It must also continue to check for any preamble bytes since memory grant could be de-asserted during the middle of preamble transmission for a fully valid frame. This prioritizes throughput, leaving a more flexible length of time for memory to stall.

Lastly, when the parsed destination address is not recognized by the switch, the flooding mechanism sends data out to every egress port. This is demonstrated in figure \ref{fig:packet_drop} where the input on port 0 gets sent to every osther port.

\section{SRAM}
The SRAM is an array of 64 registers, each of size 64 bytes for a total of 4096 bytes. It has 1 read port and 1 write port, both of which have a 1 cycle latency. The SRAM block size is 64 bytes to match the register size. Every SRAM block has 63 bytes of payload and 1 byte reserved for the footer. The footer has 6 bits corresponding to the next index, and 1 bit denoting whether this block contains the end of a packet (eop). The footer is necessary so that the read controller can traverse non-contiguous blocks using the next pointer (linked list style) until it reaches the eop flag.

\section{Free List} \label{free-list}
\begin{figure}[H]
    \centering
    \includegraphics[width=1\linewidth]{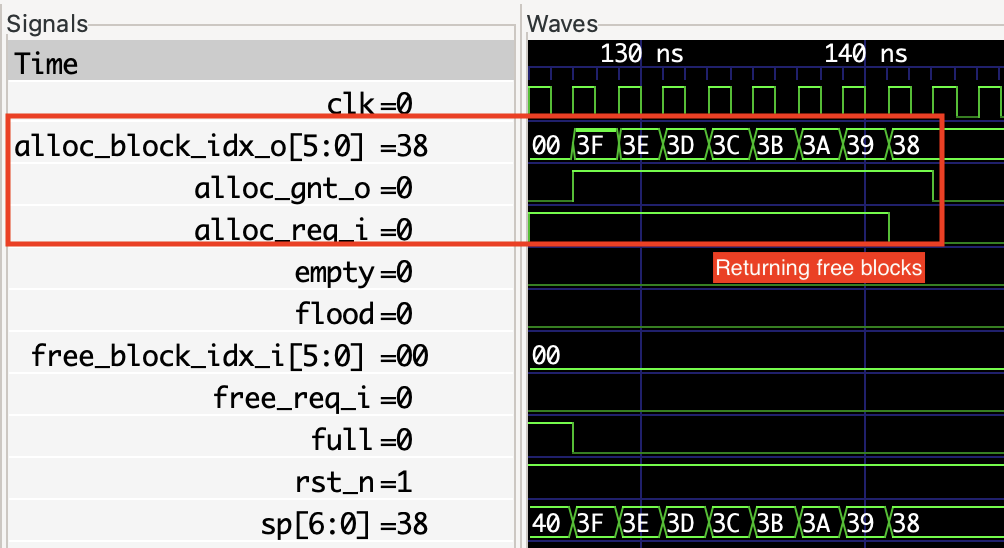}
    \caption{Allocation requests to free list}
    \label{fig:fl_waveform}
\end{figure}

The blocks in SRAM are managed by a free list, which contains all currently free blocks. It is implemented as a stack of available memory block indices. Allocating a block in memory requires popping from the free list, and freeing a block of memory pushes to the free list. The free list can handle simultaneous allocations and frees. In the event of an address table miss, a packet must be sent on all egress ports (a flood). Therefore, the associated packet memory can only be freed once the last port has transmitted the data. We accomplish this by keeping a small array of reference counters per memory block, and only return a block to the free list once the reference counter reaches 0. 

\section{Memory Write Controller}

The memory write controller interfaces with the RX module and memory. There is one memory read controller per port. Because the RX module sends 1 byte per GMII clock cycle, the memory write controller needs to coalesce these bytes into a memory block. After 63 bytes of payload has arrived, or the RX mac controller signals the end-of-frame, the memory write controller writes the footer. During the coalescing period, the memory write controller secures allocations for the current block of memory and the next block of memory, which is necessary to correctly write the next index field of the footer. This allows interactions with the free list to generally be hidden and off the critical path. The memory write controller has 4 states: IDLE, WRITE PAYLOAD, WAIT, and FOOTER. Byte data is coalesced in the WRITE PAYLOAD state. The WAIT state is optional, and it is only entered if SRAM block allocations are still pending from the free list. The footer state involves writing the footer and scheduling a memory write request. We remain in the footer state until the arbiter grants access to the memory write port. If we are currently in the WAIT state, or we are in the FOOTER state and waiting for memory write port access, the write controller pulls it's ready signal low to the RX module indicating it is currently not in a state to receive new data. These back-pressure scenarios are rare and occur in less than 3\% of cycles.

\begin{figure}[H]
    \centering
    \includegraphics[width=0.9\linewidth]{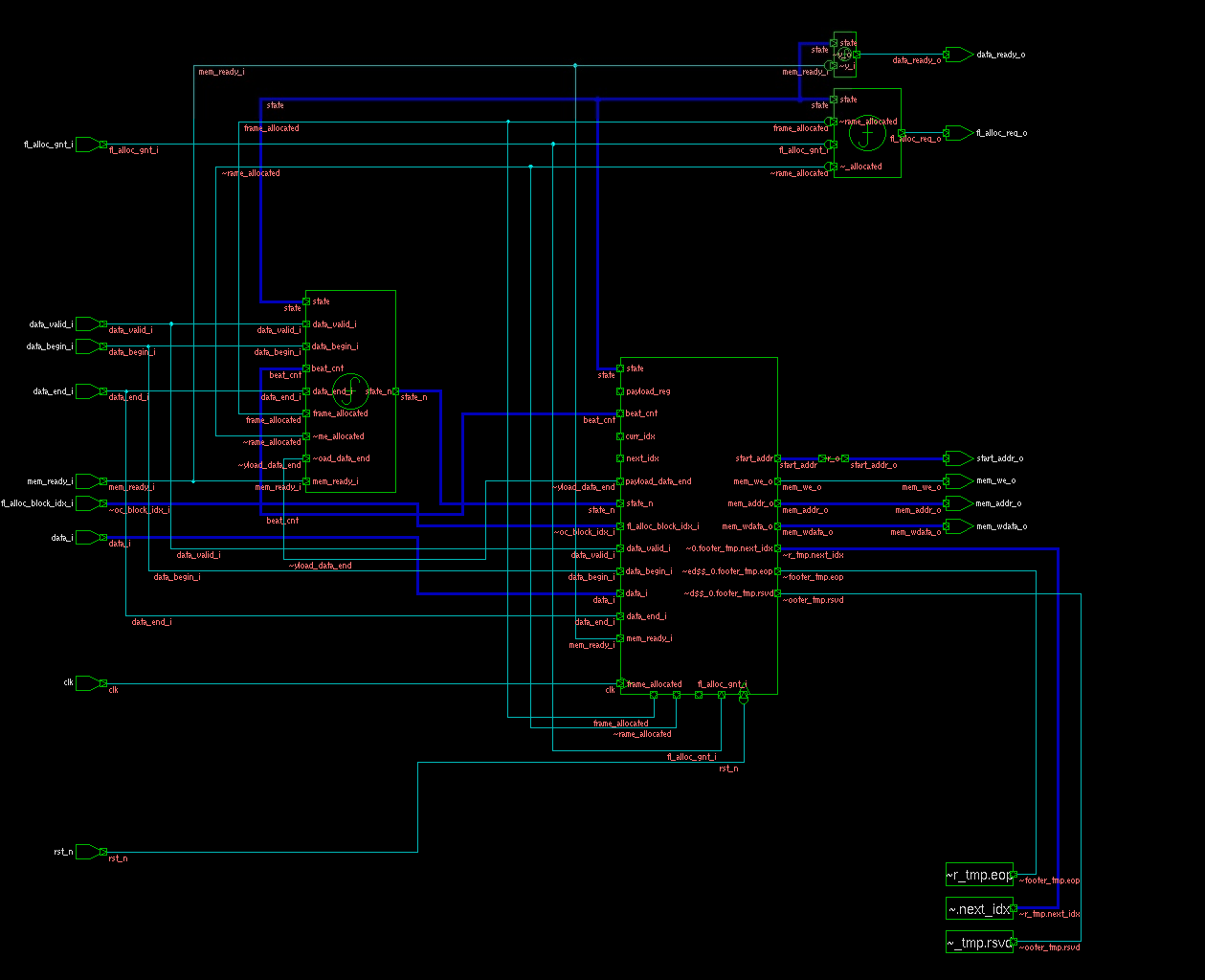}
    \caption{Memory write controller schematic.}
    \label{fig:mem_wr_ctrl}
\end{figure}

\begin{figure} [H]
    \centering
    \includegraphics[width=1\linewidth]{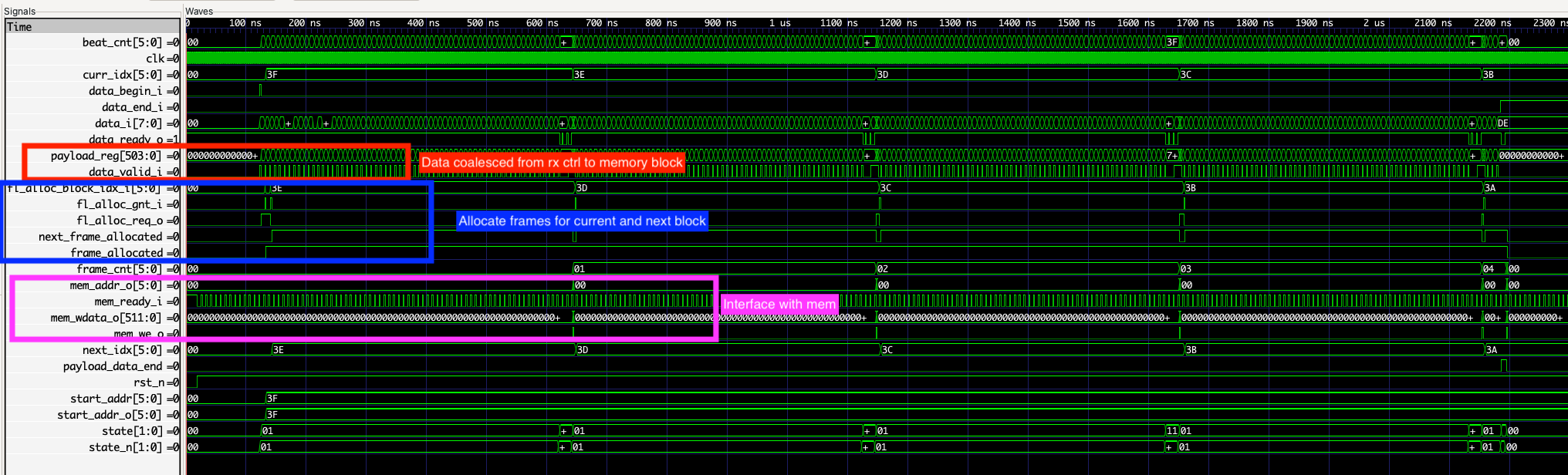}
    \caption{Memory write controller waveform.}
    \label{fig:mem_wr_ctrl_annot}
\end{figure}

\section{Memory Read Controller}
The memory read controller interfaces with the TX module and the memory read port. There is one memory read controller per egress port. The memory read controller is a simple module that traverses the linked list of blocks corresponding to the packet until it reaches the end-of-packet flag. It returns the blocks one at a time to the TX module, and frees the blocks in the process, pushing them to the free list.
\begin{figure} [H]
    \centering
    \includegraphics[width=1\linewidth]{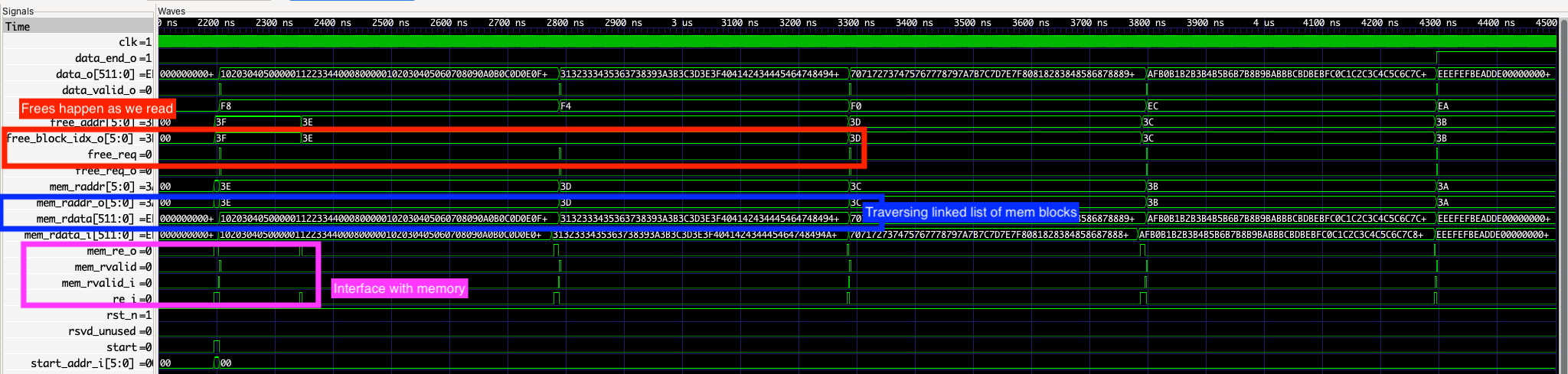}
    \caption{Memory read controller waveform.}
    \label{fig:read_ctrl_annotated}
\end{figure}
\section{Arbiter} \label{arbiter}
The arbiter manages access to the SRAM IO ports, address learn table, and free list. The per-port memory read controllers contend for access to the SRAM read port and the free list free-port (only 1 free may happen at a time). The per-port memory write controllers contend for access to the SRAM write port and the free list allocation port. The ingress and egress modules contend for access to the address learn table. For each of these shared resources, access is granted in a round-robin fashion. Memory allocations are a special case, as requests through the free list are guaranteed to be granted in order to prevent a port from getting starved. 

\section{Crossbar} \label{crossbar}

The crossbar module connects the input memory write to the correct output output port. The transmission starts upon the end of frame signal from an ingress port. Instead of sending the whole frame through the crossbar, this implementation only transmits a memory pointer through the crossbar. The output port reads from memory with the given pointer to the GMII output port. The crossbar must handle two connection cases: learned address, flooding. In order to recognize which port is connected to the destination address recipient, the crossbar maintains a fully associative cache called the address learn table. Every source address is mapped to the ingress port in the cache. 

If the destination address has been recognized from a previous transmission, the crossbar will route the memory pointer to the destination port. However, in the case of a cache miss, the crossbar will flood the data to all egress ports. This process leads to a memory contingency discussed in section \ref{free-list}. 

\begin{figure}[H]
    \centering
    \includegraphics[width=0.9\linewidth]{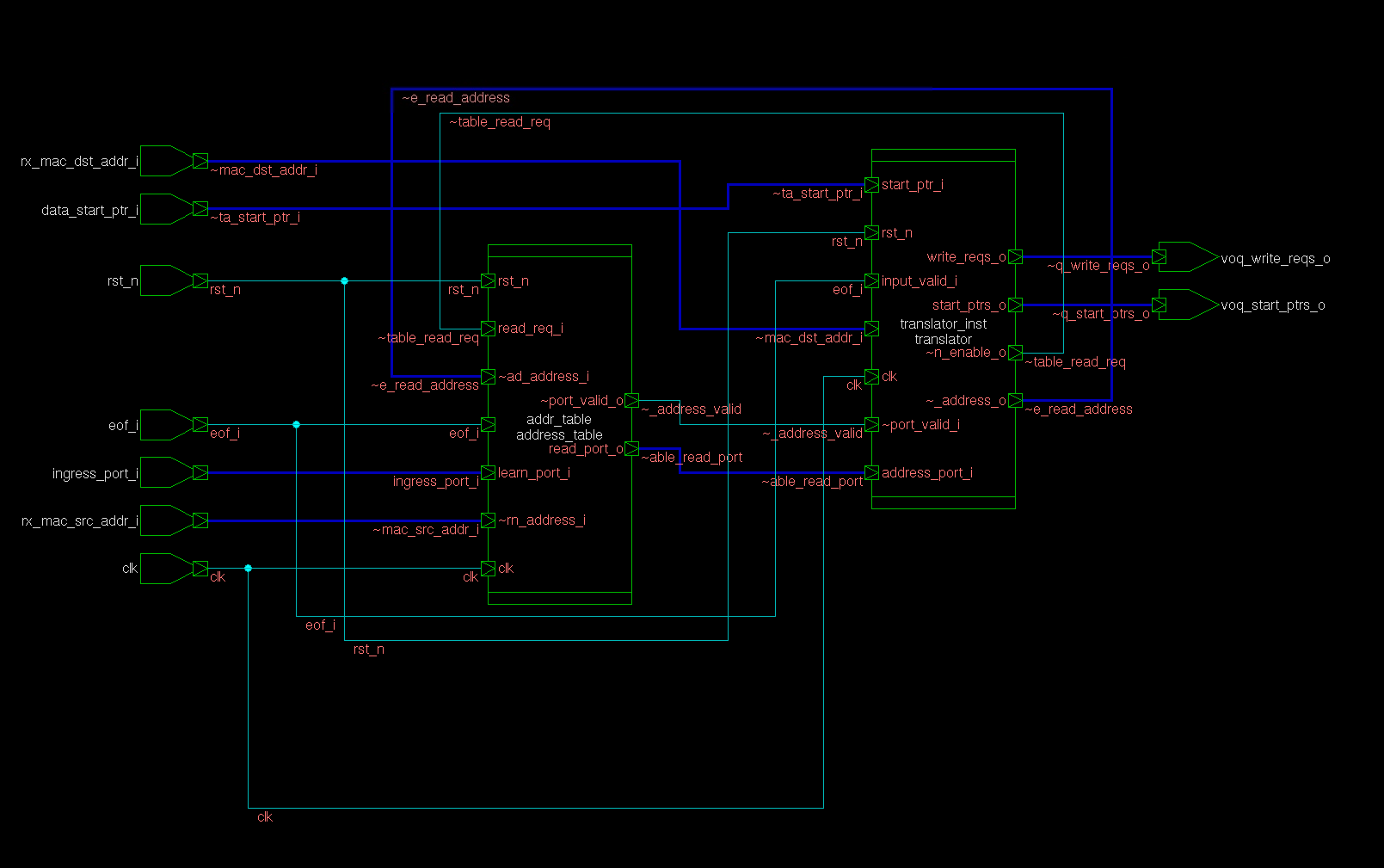}
    \caption{Crossbar Schematic}
    \label{fig:xbar_waveform}
\end{figure}

\begin{figure}[H]
    \centering
    \includegraphics[width=1\linewidth]{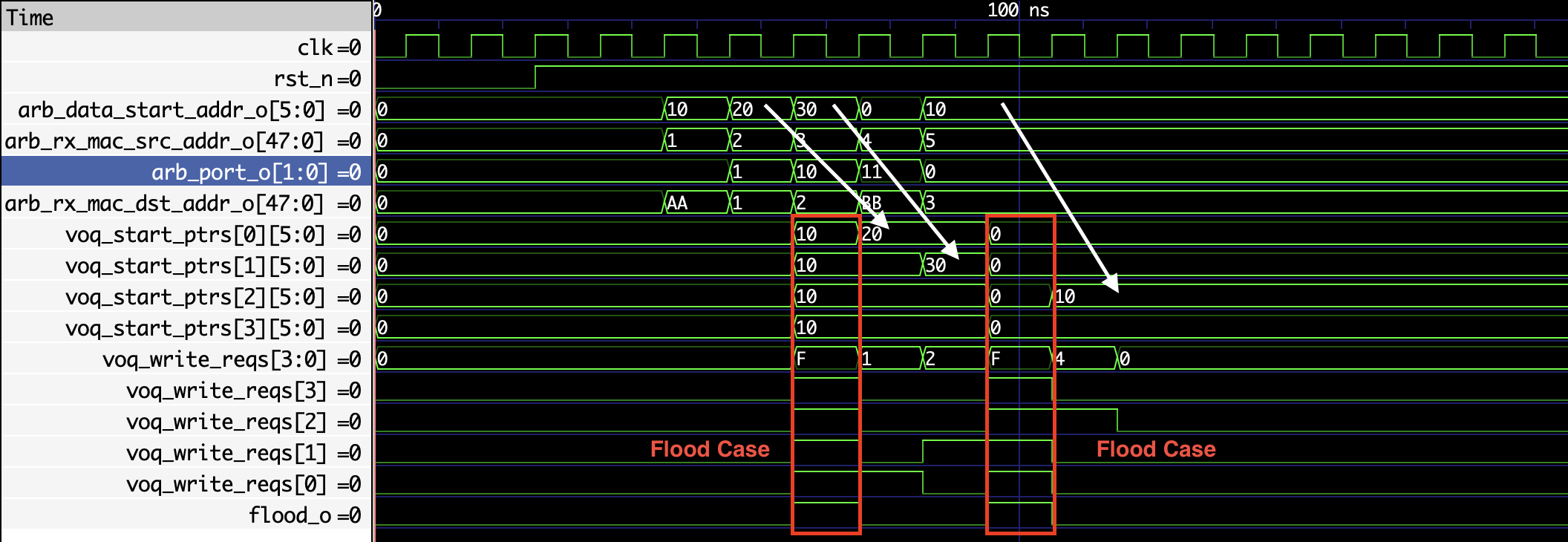}
    \caption{Crossbar Learning Testbench Waveform}
    \label{fig:crossbar-waveform}
\end{figure}

The completed crossbar functionality is verified in figure \ref{fig:crossbar-waveform}. The overall schematic can be found in figure \ref{fig:xbar_waveform}

\subsection{Address Learn Table}


The address learn table receives a source address along with a port number. The goal of the address table is to learn this information whenever the data is valid. Then, the table should output the corresponding port number for previously learned addresses.

This functionality requires arbitration of ingress ports and their source address outputs. The original address table implementation utilized the round robin scheme to learn one port at a time. Since the source address is the second part of the frame that is read in, each port can safely be evaluated once at a time before the next frame arrives. Once again, the tradeoff between multiple write ports to the address learn table are similar to those discussed in the overall architecture section \ref{overall_arch}. 

In the case of the address learn table, implementing the global arbiter handled any necessary arbitration. With the consolidation of round robin schemes into one module described in section \ref{arbiter}, the address learn table could be simplified. Instead of going around each port input, now, if the single end of frame signal from the arbiter is high, the cache maps the input source address to the active port.

Another requirement of the address learn table is eviction. The industry standard for this is to use a five minute age out policy. For this implementation, we arrived at a pseudo least recently used replacement policy. The reason for this modification is because this Ethernet switch implementation does not rely on any external intellectual properties. Therefore, all memory modules, including the SRAM, are small in size. The address table only maintains 16 entries. To ensure that addresses continue to learn even after the small cache is full, we utilize a cache eviction policy instead of the standard age out mechanisms.

The cache eviction policy the address learn table uses is maintaining a hit counter on each row of the table. Every time an entry is read out, the counter for that element is incremented. At the same time, all other counters are decremented. To ensure that a new element isn't immediately evicted, all counters are initialized at a value of 1. Therefore, if an address mapping isn't utilized, then it will have the lowest minimum hit value and be replaced.

When the address table was first synthesized, the slack time was negative. This was fixed in a couple of different ways. The first attempt was by simply reducing the number of entries. This minimized the combinational logic on the eviction policy. However, the entries had to be reduced to one per port to satisfy timing constraints. This number was too small, so we ran another attempt by lowering the number of bits on the counter. In this case, using two bits was enough to meet timing constraints. Finally, by running the synthesis at Ma\&P set to high effort, the slack achieved 0. This is further discussed in \ref{synth_flow}.

\begin{figure} [H]
    \centering
    \includegraphics[width=1\linewidth]{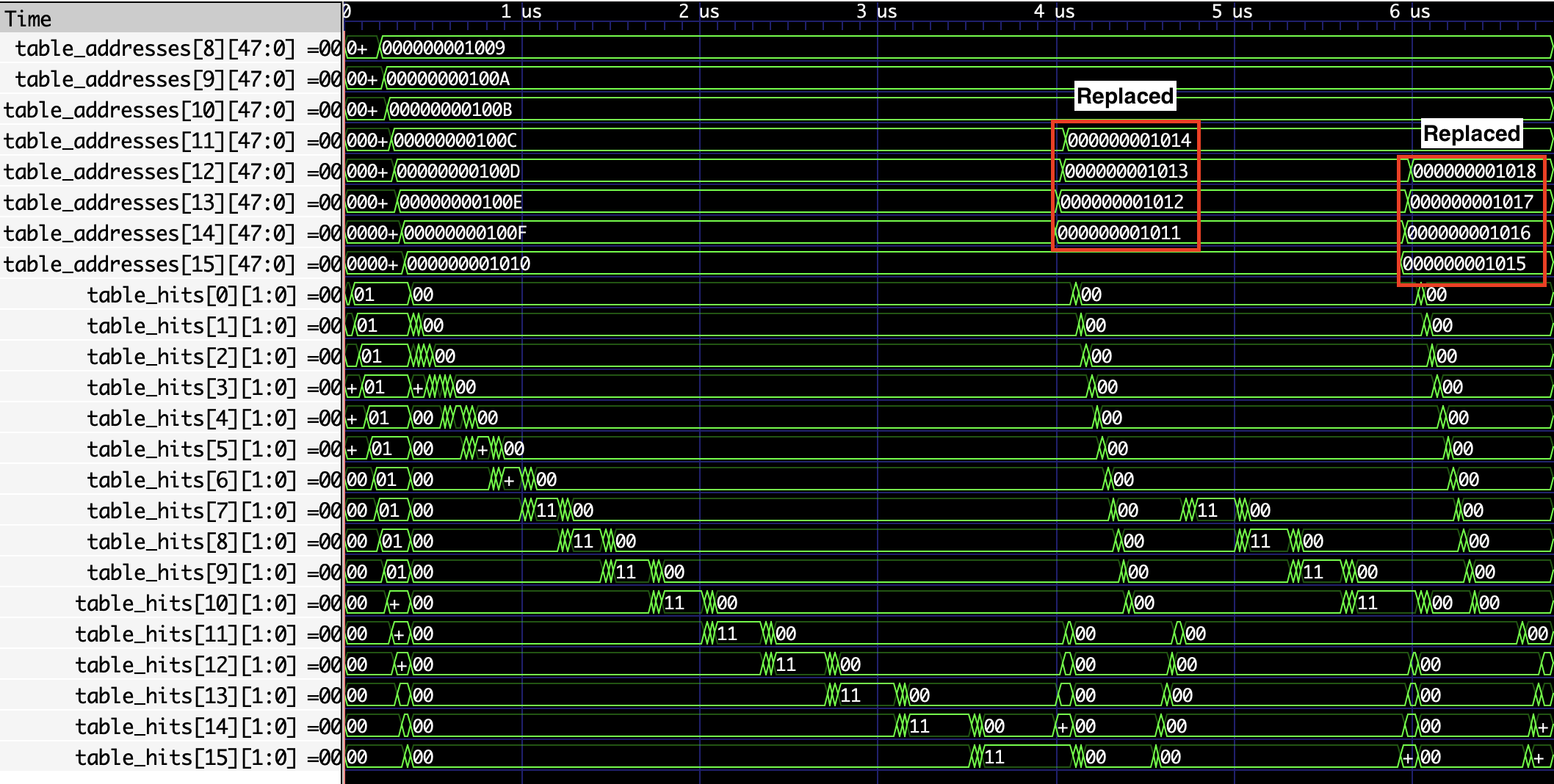}
    \caption{Address Learn Table Cache Eviction Test}
    \label{fig:address_learn_table_wave}
\end{figure}

\subsection{Router}

The router module takes in the memory start pointer of the ingress frame and pushes it into the virtual output queue of the corresponding egress port. The inputs of the router are the destination address and the start pointer. The module sends a request to the address learn table and routes the data on the next clock cycle. If the address learn table does not output a valid port number the router defaults to a flood signal output. This means that the start address is sent to all the VOQ buffers. Additionally, there is a flood tag sent with the data. This way, when the egress port reads from memory, the index will only be freed when all ports have read the flood data. This is further discussed in section \ref{free-list}.

\section{Virtual Output Queue}

The Virtual Output Queue (VOQ) stores the addresses in memory for output frames. Additionally, it also stores a corresponding bit with each frame that indicates whether it is a flood frame. This flood logic is implemented in section \ref{free-list}. The VOQ also supports same cycle read and writes on a full queue as well as same cycle read and writes on an empty queue. These two features maximize throughput from the crossbar to the transmitting module. The enable on the pop functionality the ready signal from the egress port described in section \ref{egress-port}. The enable on the push for the queue is the voq write signal from the crossbar described in section \ref{crossbar}. The functionality is verified in the following waveform. The input addresses are outputed in order when the read signal is triggered.

\begin{figure} [H]
    \centering
    \includegraphics[width=1\linewidth]{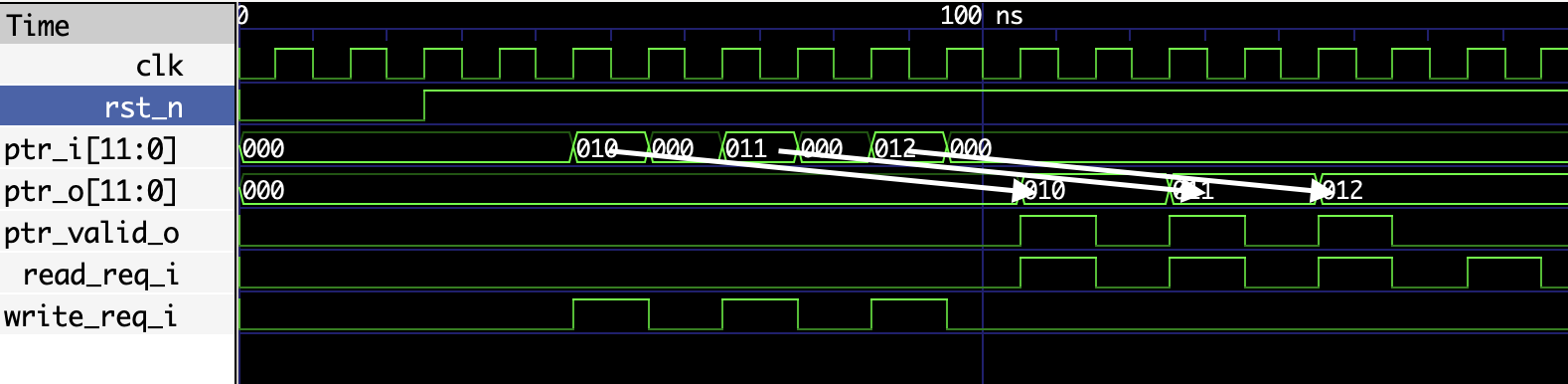}
    \caption{VOQ functionality test waveform}
    \label{fig:voq_waveform}
\end{figure}

\section{Egress Port} \label{egress-port}

\begin{figure} [H]
    \centering
    \includegraphics[width=.75\linewidth]{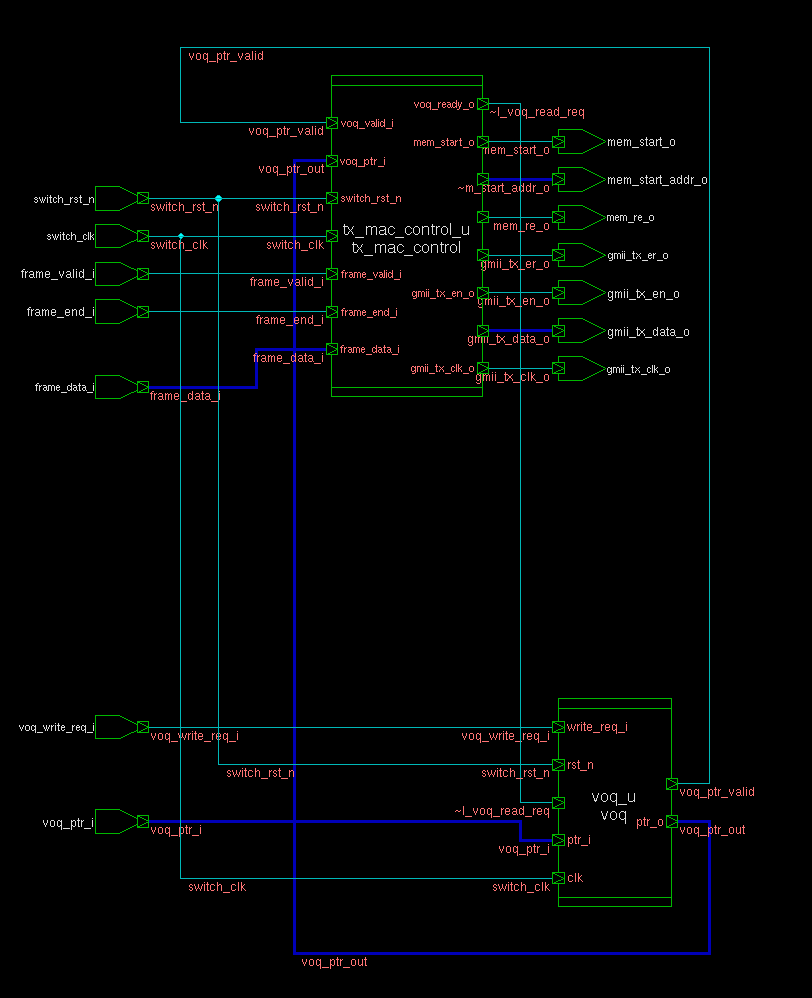}
    \caption{Egress module schematic}
    \label{fig:egress_port}
\end{figure}

At the egress of each port is the TX module. This interfaces between each VOQ and memory read controller with each port's respective GMII transmit output signals. The TX takes begins by waiting for the starting address pointer from the VOQ. TX handshakes with VOQ using a VOQ ready bit that is always high when the TX is ready to start a new frame. After waiting for a VOQ valid bit to indicate that the VOQ has a valid starting pointer, the VOQ ready is pulled low and the VOQ starting pointer is passed to the memory read controller.

\begin{figure}[H]
    \centering
    \includegraphics[width=1\linewidth]{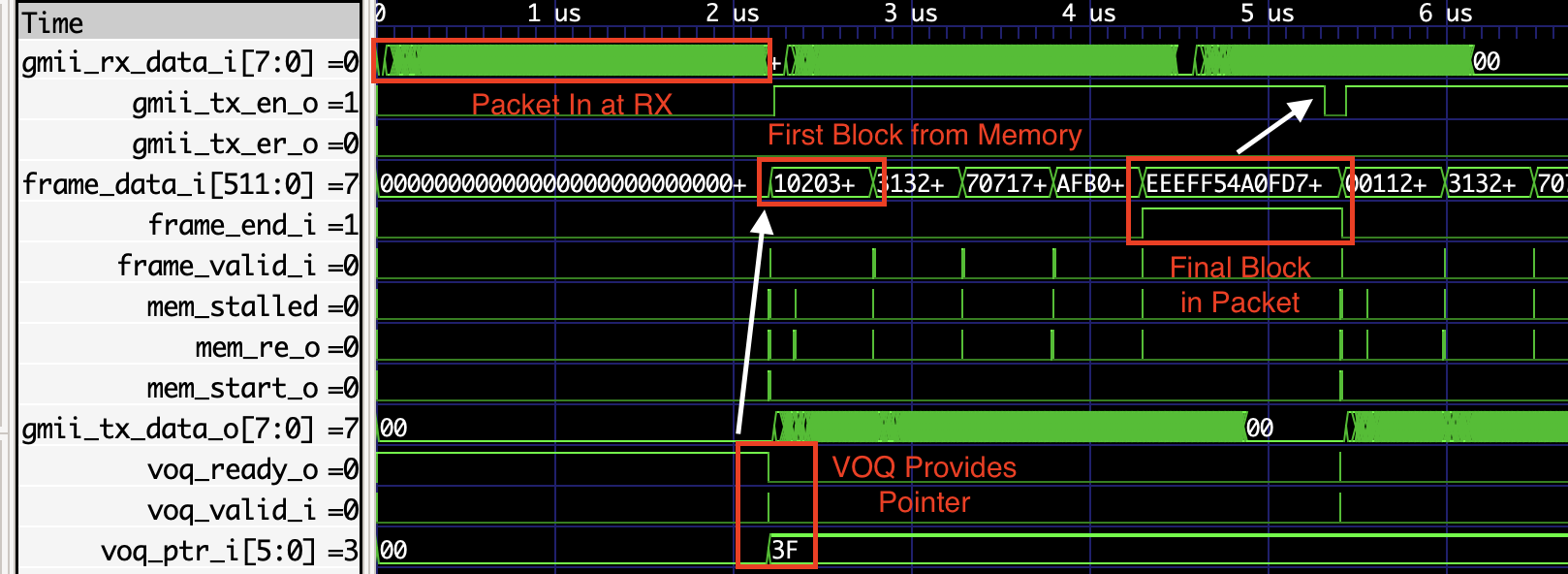}
    \caption{Packet drop functionality with corrupted CRC}
    \label{fig:tx_mac_control}
\end{figure}

As seen in figure \ref{fig:tx_mac_control} when beginning the frame transmission, the TX module first ensures that memory read is not stalled. A data packet comes in at RX. Then, TX checks for the VOQs starting pointer. Once this is given, memory can provide the frame data to TX, and sets the frame end on the last block of the packet of data for TX to handle. If stalled, TX continues to send the frame’s starting block address pointer and requests that block from memory. Once memory read control gives back a valid full memory block, TX then moves on to sending the preamble bytes. An asynchronous FIFO is again used to interface across the CDC boundary to the GMII signals. 

At first, we prioritized throughput by jumping immediately to sending the preamble as soon as VOQ provides a valid pointer. However, all ethernet frames must be sent in contiguous bytes on the GMII clock (note, this is not true in the switch clock since the switch is 4x faster than the GMII). Otherwise, the respective receiver would have no way to know if data valid going low signifies a stall or the end of a frame. Since memory could potentially stall for longer than the full preamble transmission time, this could lead to an undesirable result where the current frame being transmitted must be flushed out and preamble must be resent when memory is fully ready. Thus, waiting to send preamble prioritizes robustness over full-on performance, choosing a more reliable waiting method over a speculative jump ahead to sending the preamble.

\section{RTL to GDSII Flow} \label{synth_flow}

The synthesis flow for this project utilizes the \texttt{freepdk45} library. Since the goal of the Ethernet switch implementation was primarily on a functionality level, we analyzed PPA from an architecture tradeoff perspective. Therefore, utilizing the aforementioned mentioned dummy was enough for our purposes. 

Design vision was used to synthesize the netlist for the APR. In design vision, apart from the \texttt{freepdk45} library selection, the wire load configuration was set to none. Once again, since the purpose of the implementation was to determine architecture level tradeoffs, adding a wire load does not provide benefits in synthesis. 

Finally, five clocks are specified in the synthesis flow. The five clocks are the switch clock and a separate gmii rx clock for each ingress port. The frequency of the switch clock is four times the gmii rx clock to maximize throughput and minimize packet losses from back pressure. Since the gmii frequency is set by the protocol requirements, the switch clock frequency is 500 MHz. 

When compiling the design, the performance constraint effort is set to high. This ensures that the slack time is not negative. Since several modules barely meet timing closure, placing the synthesis on high effort is necessary.

After the synthesis is complete, we ran an Automatic Place and Route on Cadence Innovus. The APR was primarily to extract the parastics from the design. This is fed into PrimeTime to verify the timing closure after the place and route is complete.

\section{Results}

\begin{figure} [H]
    \centering
    \includegraphics[width=1\linewidth]{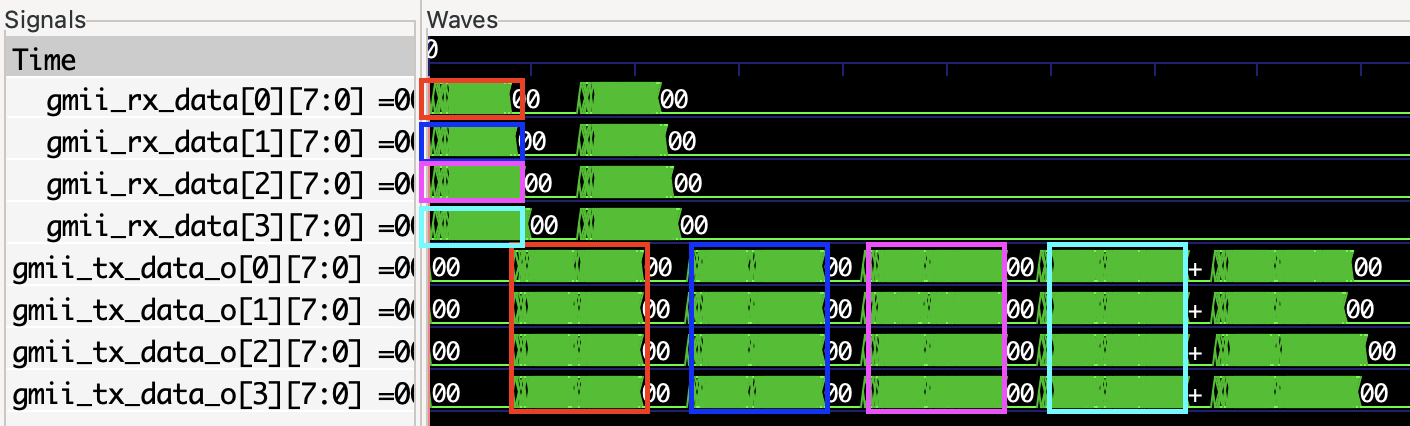}
    \caption{First packet on each port floods.}
    \label{fig:multiport_overall}
\end{figure}

Our discussion of the modules above refer to the final implementations. We will discuss some design iterations and their results in this section. Figure \ref{fig:multiport_overall} shows the result of a test bench that sends two packets on each port in parallel. The first packet sent on each port is flooded, since the mapping between destination address and port has not been learned yet.

\begin{figure} [H]
    \centering
    \includegraphics[width=1\linewidth]{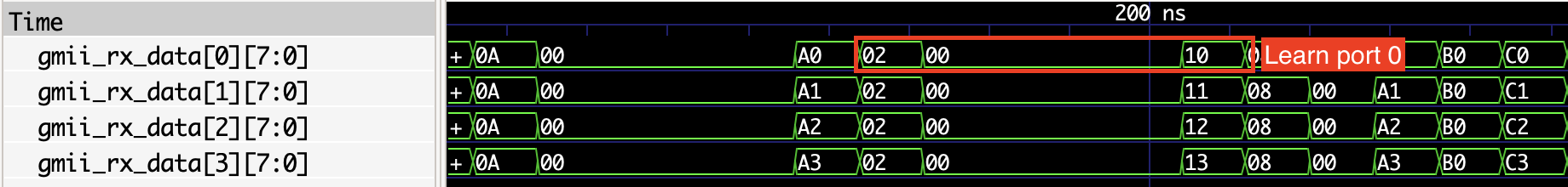}
    \includegraphics[width=1\linewidth]{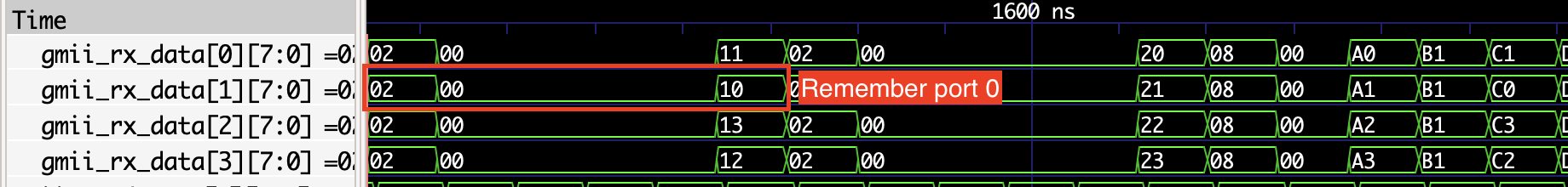}
    \includegraphics[width=1\linewidth]{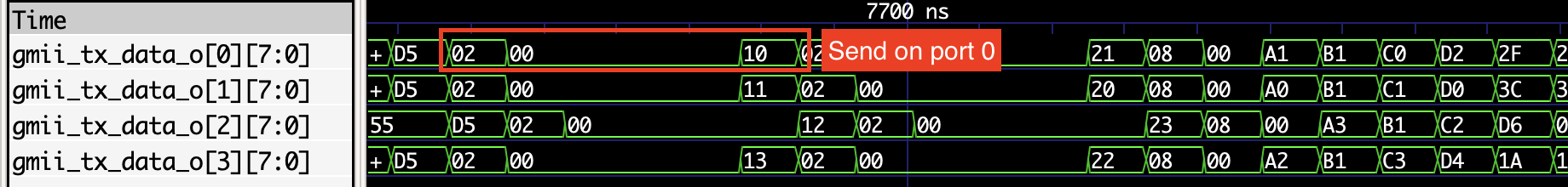}
    \caption{Packets are routed to correct egress ports with no data corruption.}
    \label{fig:multiport_waveforms}
\end{figure}

The second set of packets have a destination address that matches one of the source addresses in the first set of packets. This allows us to test the address learning functionality. As can be seen in the waveform, these packets are routed to the correct output port with no data loss.

\subsection{Base Priority Encoder Free List Allocation Scheme}
An early iteration of the free list was a brute-force priority encoder to select the next free block. We expected the power cost to be a bit high in this case, but performance potentially better. In reality, power, performance, and area were negligibly impacted when we upgraded to the stack plus reference counter based free list implementation. This base priority encoder free list had a correctness flaw, however. If in the flooding case the floods for a particular packet were not exactly in lock step, we could potentially free a block early, leading to incorrect behavior. Our upgraded final design fixes this error. 
\begin{figure} [H]
    \centering
    \includegraphics[width=0.75\linewidth]{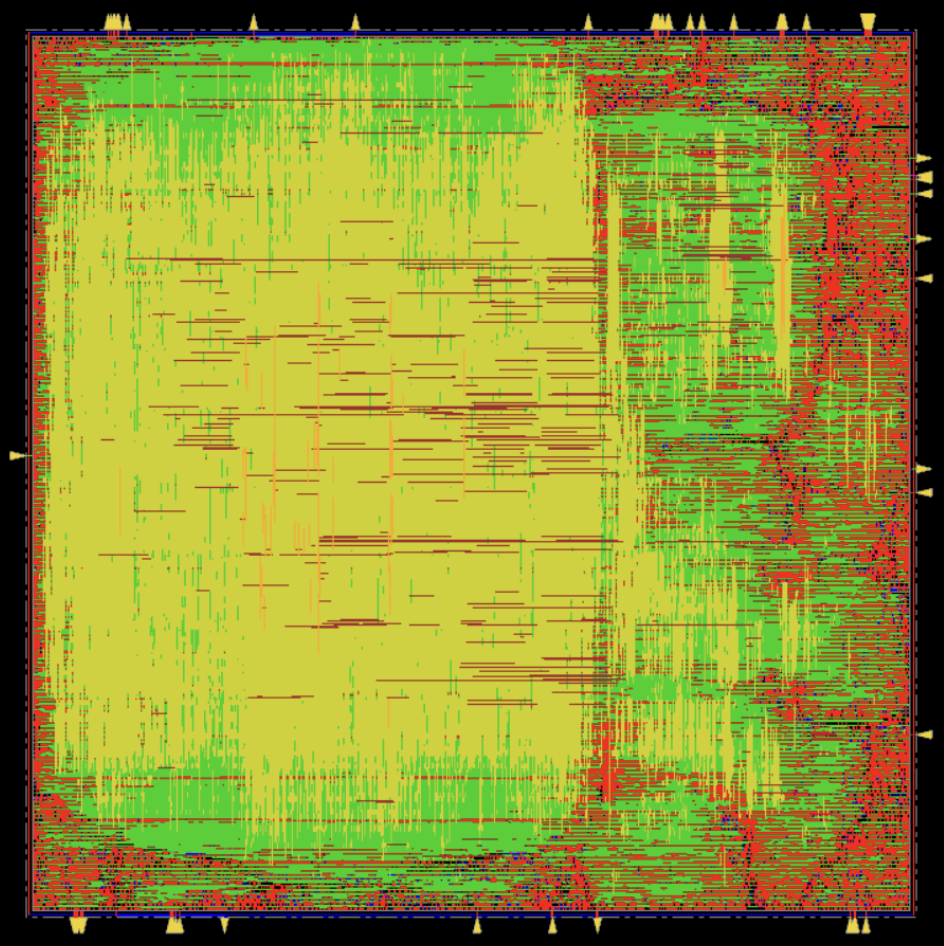}
    \caption{APR With Base Priority Encoder FL Run}
    \label{fig:APR_Base}
\end{figure}

\subsection{Free List Flood Counter Run}

\begin{figure} [H]
    \centering
    \includegraphics[width=0.75\linewidth]{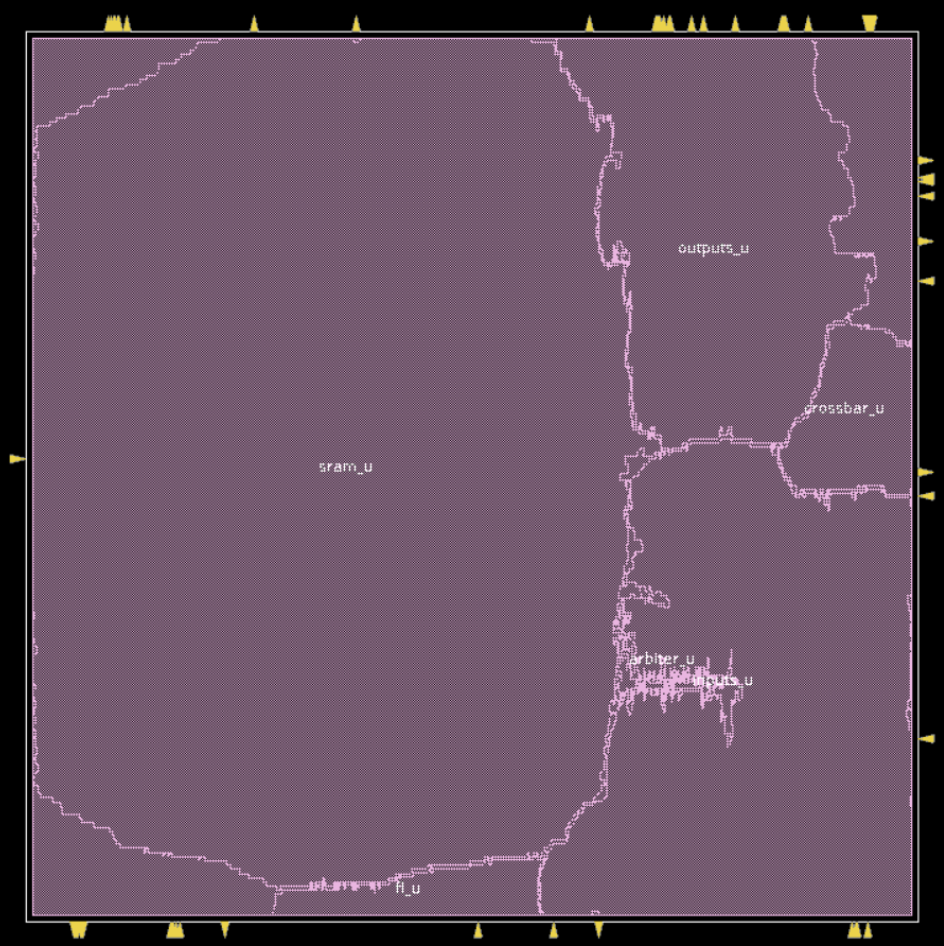}
    \caption{Amoeba View of the Free List Flood Counter Enabled Run}
    \label{fig:Amoeba_View}
\end{figure}

The implementation that is discussed in the above modules utilizes a stack based free list. This is computationally less intensive than a priority encoder. However, to handle the flood case, there is an array that counts the number of reads before freeing. This implementation is discussed in \ref{free-list}. The following APR layouts show the congestion and placement of different modules throughout the Ethernet switch chip. From figure \ref{fig:Amoeba_View}, we can tell that SRAM, as expected, takes up the most area. In the future, utilizing an external IP would allow for the physical layout to be smaller and more efficient. In fact, the post APR slack time, as discussed in this report does not meet timing closure because of the SRAM delay. In the future, to minimize this, we would use a banked external memory, which has a lower delay. Additionally, as we discussed further, a pipelined memory could allow for minimizing the effect of SRAM on the overall slack.

\begin{figure} [H]
    \centering
    \includegraphics[width=0.75\linewidth]{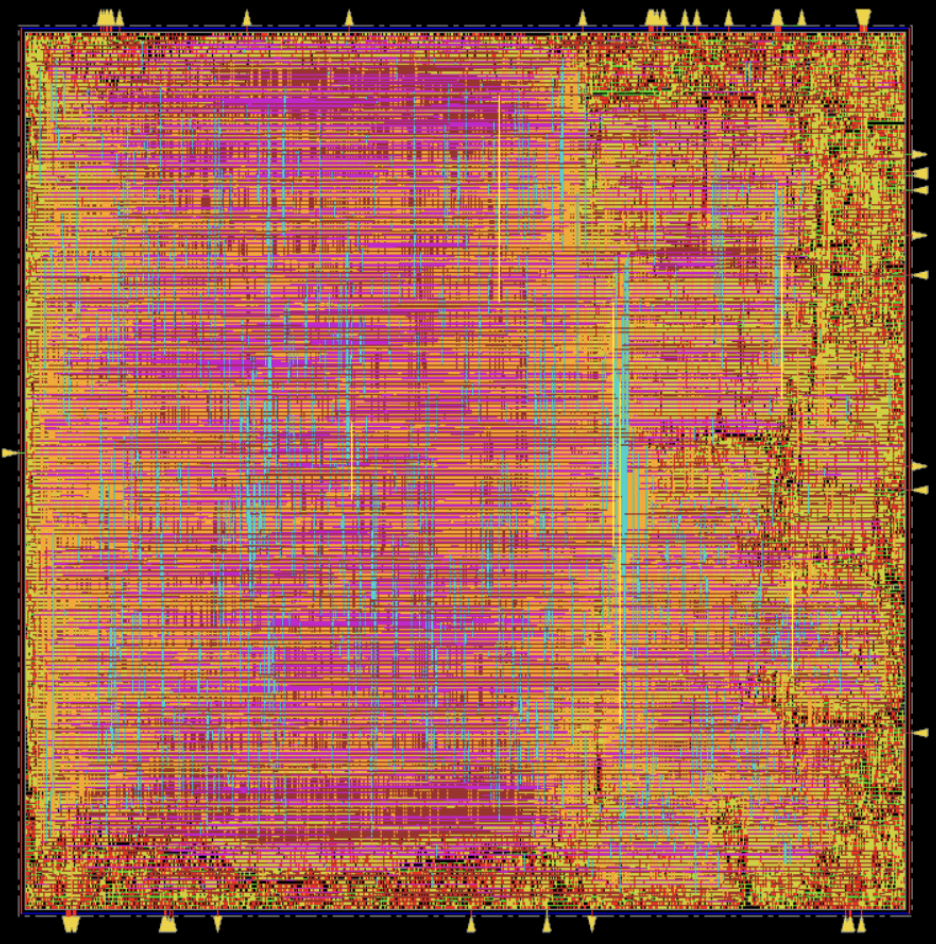}
    \caption{Floorplan of the Free List Flood Counter Enabled Run}
    \label{fig:Floorplan_Flood}
\end{figure}

\begin{figure} [H]
    \centering
    \includegraphics[width=0.75\linewidth]{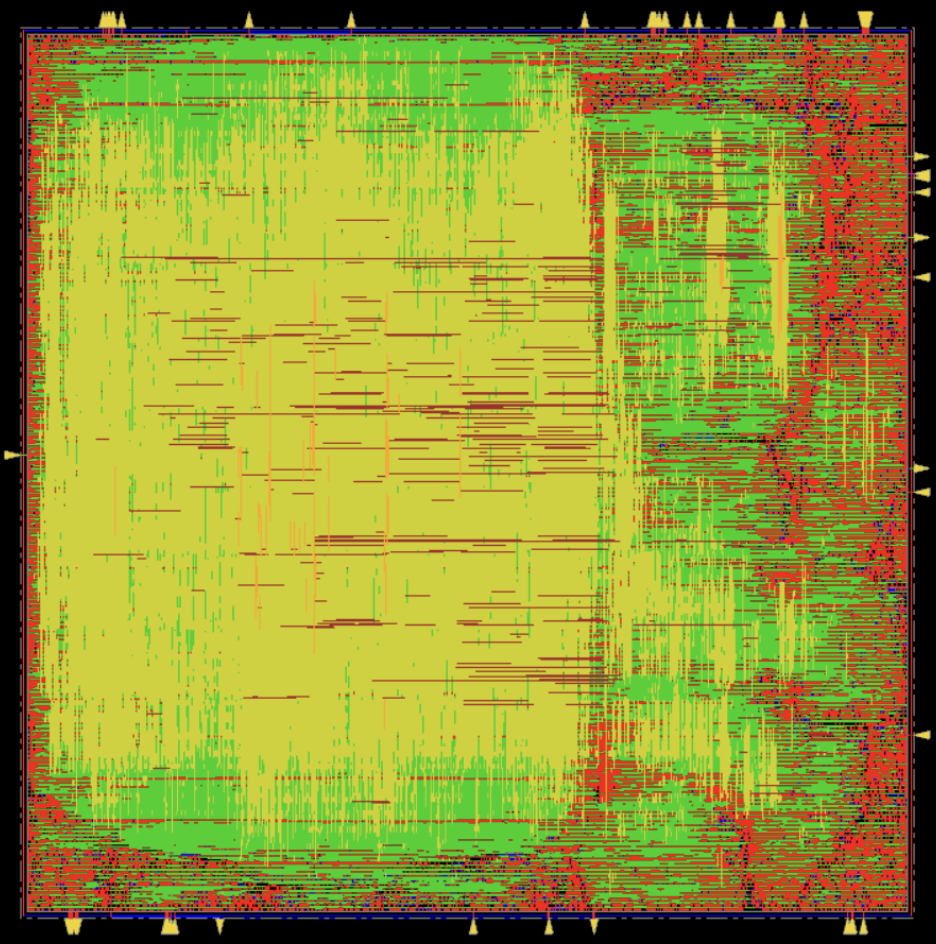}
    \caption{Full APR the Free List Flood Counter Enabled Run}
    \label{fig:APR_Flood}
\end{figure}

\subsection{Area, Timing, and Power}

In table \ref{tab:performance} running Design Vision, we were able to produce the following results for area, timing, and power for the two designs. Notice that the numbers have a negligible difference, showing that our design choice of using the priority encoder free list versus the stack-based free list shows minimal difference from a PPA perspective. We can see that slack has been optimized to exactly 0ns, meeting all timing requirements while minimizing area and power as much as possible.

\begin{table}[H] \scriptsize
    \caption{Design Vision Synthesis Runs}
    \centering
    \begin{tabular}{|c|c|>{\raggedright\arraybackslash}p{1.5cm}|}
        \hline
         \rowcolor{lightgray} \textbf{Pin Name} & \textbf{Base Design}& \centering\arraybackslash\textbf{Flood Design}\\
         \hline
         Total Cell Area (um\^2)& 1,014,550& 1,014,964\\
         \hline
         Critical Path (ns)& 1.91& 1.92\\
         \hline
         Slack (ns)& 0.00& 0.00\\
         \hline
         Internal Power (mW)& 379.67& 379.67\\
         \hline
         Switching Power (mW)& 3.07& 3.06\\
         \hline
         Total power (mW)& 388.13& 388.12\\
         \hline
    \end{tabular}
    \label{tab:performance}
\end{table}
As seen in table \ref{tab:performance} slack was met in design vision, and very nearly met in PreAPR, but there was significant negative slack in PostAPR. The critical path in Design Vision, PreAPR, and PostAPR consisted of the memory accesses out of the arbiter to SRAM, which was unsurprising. We modeled SRAM access as one cycle in RTL for simplicity, but this clearly did not meet timing once physical constraints came into play. There are significant decoding costs and distance costs to interact with memory, which prevents it from meeting timing in a single cycle. The obvious solution would be to pipeline the memory access, which could easily be implemented given that we support ready-valid handshakes at every interface. 
\begin{table}[H] \scriptsize
    \caption{APR Timing (PrimeTime)}
    \centering
    \begin{tabular}{|c|c|>{\raggedright\arraybackslash}p{1.5cm}|}
        \hline
         \rowcolor{lightgray} & \textbf{PreAPR}& \centering\arraybackslash\textbf{PostAPR}\\
         \hline
         Slack (ns)& -0.01& -16.03\\\hline
    \end{tabular}
    \label{tab:postapr}
\end{table}

\section{Conclusion and Future Work}

The L2 Ethernet Network Switch implementation discussed in this report meets the goal of addressing compliance with the IEEE 802.3 Ethernet standard. Along with being one of the few open source implementations, it also reveals many of the different architectural tradeoffs that are possible within the implementation. In the future, some modifications could be made to increase throughput and minimize inefficiencies. The goal of this paper was accomplished, as it reveals many of those alternatives.

One of the main options for future development is the memory. As discussed earlier, memory is part of the reason that the post APR simulations do not pass timing constraints. Utilizing a separate banked memory IP and integrating a pipelined architecture would relieve some of the pressure on the timing constraint. Apart from this change, adding more memory read and write ports would allow for less arbitration and a lower clock speed for maximum through put. Testing this option against the alternatives outlined in the paper would allow for a more comprehensive Ethernet Switch analysis.

The address learn table is another module that has opportunity for improvement. Once again, this is a module that has a critical path that barely meets timing constraints. To fix this, the eviction policy could be updated to an age out system. By using external memory IP, the address table cache depth could be larger. This would allow for an age out system instead of an eviction scheme. The change would also expire old addresses, which may no longer be found on a specific output port.

One of the only edge cases that isn't completely solved in the implementation is the flood case where some of the VOQs are full. Since a flood tag means the memory waits for all ports to have read the data frame, if a VOQ drops a packet because it is full, the memory will never free the flood frame index. Changing this would allow for flooding to work in all cases and prevent memory from filling up.

The last IP that could be utilized to strengthen the APR is a clock tree. The clock currently has a fanout higher than 4000 connections. This means that the logical effort on this node is really, and the clock source may experience driving delays. Utilizing a clock tree would minimize unintended skews in the design. 

Apart from these changes, the project can be improved by making memory reads more secure. Networking has several contingencies, of which we only address a small subset. Every decision has a large impact on the PPA performance of the chip. Investigating more implementations will only reveal further opportunities for improvement.

\bibliographystyle{unsrt}
\bibliography{refs}

\end{document}